\documentclass[5p,twocolumn,times,number]{elsarticle}

\usepackage{graphicx}
\usepackage{amsmath}   

\begin{document}

\begin{frontmatter}

\title{ Design and status of the Mu2e electromagnetic calorimeter}

\author[a]{N.~Atanov}
\author[a]{V.~Baranov}
\author[a]{J.~Budagov}
\author[e]{R.~Carosi}
\author[e]{F.~Cervelli}
\author[b]{F.~Colao}
\author[b]{M.~Cordelli}
\author[b]{G.~Corradi}
\author[b]{E.~Dan\'e}
\author[a]{Yu.I.~Davydov}
\author[e]{S.~Di Falco}
\author[e,g]{S.~Donati}
\author[b,l]{R.~Donghia}
\author[c]{B.~Echenard}
\author[c]{K.~Flood}
\author[b]{S.~Giovannella}
\author[a]{V.~Glagolev}
\author[i]{F.~Grancagnolo}
\author[b]{F.~Happacher}
\author[c]{D.G.~Hitlin}
\author[b,d]{M.~Martini}
\author[b]{S.~Miscetti\corref{cor}}
\ead{Stefano.Miscetti@lnf.infn.it}
\author[c]{T.~Miyashita}
\author[e,f]{L.~Morescalchi}
\author[h]{P.~Murat}
\author[e,g]{D.~Pasciuto}
\author[e,g]{G.~Pezzullo}
\author[c]{F.~Porter}
\author[b]{A.~Saputi}
\author[b]{I.~Sarra}
\author[b]{S.R.~Soleti}
\author[e]{F.~Spinella}
\author[i]{G.~Tassielli}
\author[a]{V.~Tereshchenko}
\author[a]{Z.~Usubov}
\author[c]{R.Y.~Zhu}

\cortext[cor]{Corresponding author}

\address[a]{Joint Institute for Nuclear Research, Dubna, Russia}
\address[b]{Laboratori Nazionali di Frascati dell'INFN, Frascati, Italy}
\address[c]{California Institute of Technology, Pasadena, United States}
\address[d]{Universit\`a ``Guglielmo Marconi'', Roma, Italy}
\address[e]{INFN Sezione di Pisa, Pisa, Italy}
\address[f]{Dipartimento di Fisica dell'Universit\`a di Siena, Siena, Italy}
\address[g]{Dipartimento di Fisica dell'Universit\`a di Pisa, Pisa,
  Italy}
\address[h]{Fermi National Laboratory, Batavia, Illinois, USA}
\address[i]{INFN Sezione di Lecce, Lecce, Italy}
\address[l]{Universit\`a degli studi Roma Tre, Roma, Italy}


\begin{abstract}
The Mu2e experiment at Fermilab aims at measuring the neutrinoless
conversion of a negative muon into an electron and reach a single
event sensitivity of $2.5\times10^{-17}$ after three years of data
taking. 
The monoenergetic electron produced in the final
state, 
is detected by a high precision tracker and a crystal
calorimeter, all embedded in a large superconducting solenoid (SD)
 surrounded by a cosmic ray veto system.  The calorimeter is complementary to the
tracker, allowing an independent trigger and powerful particle
identification, while 
seeding the track reconstruction and contributing to 
remove background tracks mimicking the signal.  In order
to match these requirements, the calorimeter should have an energy
resolution of O(5)\% and a time resolution better than 500 ps at 100
MeV. The baseline solution is a calorimeter composed of two disks of
BaF$_2$ crystals 
read by UV extended, solar blind, Avalanche
Photodiode (APDs), which are under development from a  JPL, Caltech, RMD
consortium. In this paper, 
the calorimeter design, the
R\&D 
studies carried out so far and the status of engineering are described.
A backup alternative setup consisting of a pure CsI crystal matrix
read by UV extended Hamamatsu MPPC's is also presented.
\end{abstract}



\end{frontmatter}

\section{Introduction}
The muon to electron conversion is an example of a Charged Lepton
Flavor Violating (CLFV) process. As all CLFV processes, it is strongly suppressed in
the Standard Model (SM), but many models of physics behind SM
predict a branching ratio accessible at current or next-generation
experiments. The conversion process is 
complementary to CLFV processes 
such as $\mu \to e \gamma$ or $\mu \to 3 e$ that reach different
sensitivity for different classes of models,
according to the relevance of loop terms or contact terms in their general  Lagrangian.
The Mu2e experiment \cite{TDR} at Fermilab is designed to reach a single event sensitivity 
of $2.5 \times$~10$^{-17}$  in the $\mu^- + Al \to  e^{-}  \; Al$
process, with an improvement of four orders of magnitude over the
current best  experimental limit~\cite{SINDRUM}. The conversion 
results in the production of a monoenergetic conversion electron, CE, with energy  equal  to the muon
rest mass apart from corrections for the nuclear 
recoil and muon binding energy  (E$_{ce}$~$=$~104.97~MeV).
Among the potential backgrounds, the muon \emph{Decay-In-Orbit} (or
DIO) is of particular concern,
since this process, $\mu^- \;
Al \to e^- \; Al \; \nu_{\mu} \; \overline{\nu}_e$,
can produce an outgoing electron that can mimic the signal in the limit
that the neutrinos have zero energy. 

\section{The Mu2e electromagnetic calorimeter}

\begin{figure}[!t]
\centering
\includegraphics[width=0.99\linewidth]{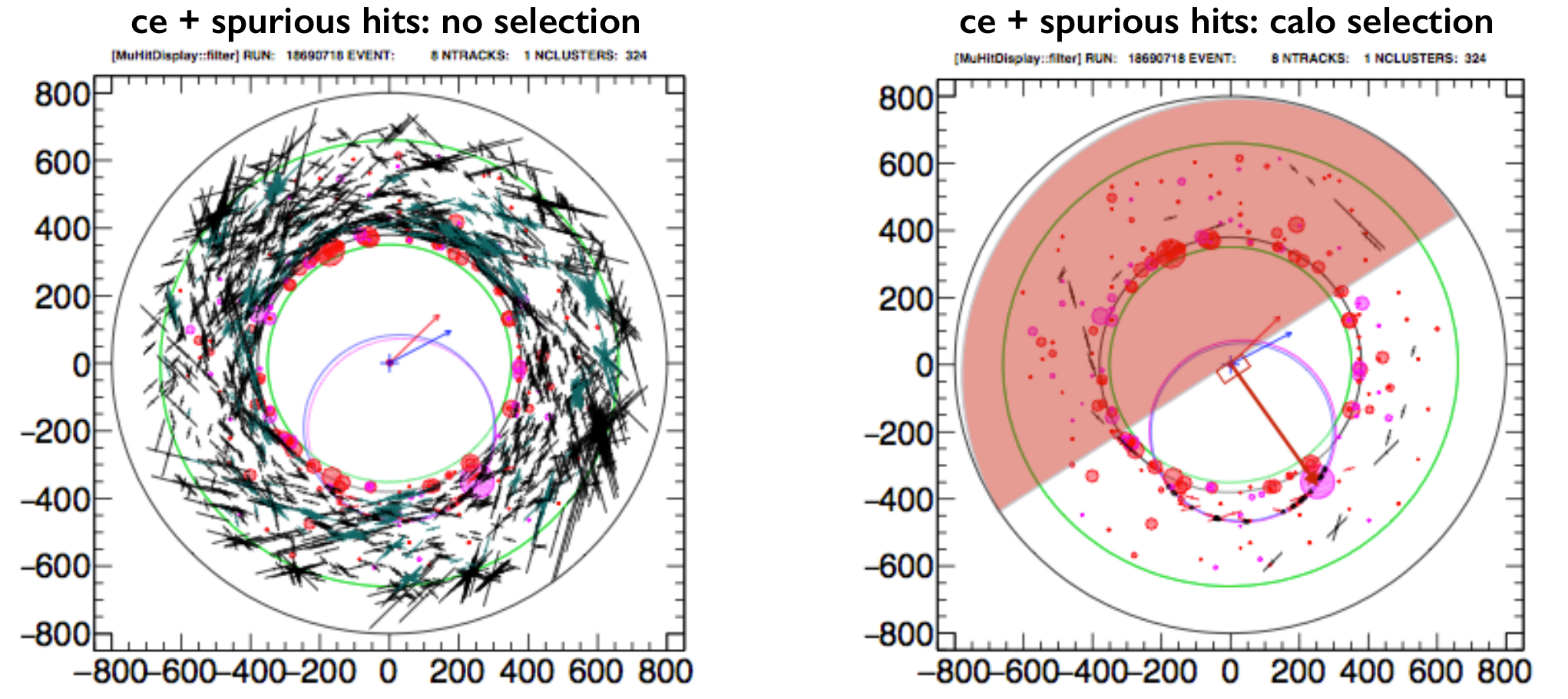}
\caption{Example of a full simulated CE event in overlap with all hits
from the environmental background: (left) without any requirement on
the calorimeter system and (right) with a calorimeter based selection.
Black points are hits from the tracker. Red points are from
calorimeter clusters. }
\label{Fig:CALTRACK}
\end{figure}

A redundant, high-precision apparatus, inside a large 
solenoid (DS), covered by a highly efficient cosmic ray veto system, 
is required to identify conversion electrons and to reduce 
background sources to  negligible level.
A tracking system, made of low mass straw tubes, provides the  high
resolution spectrometer necessary to separate signal from
background. The tracker is followed by a calorimeter system characterized by 
(i)  a large acceptance for CE's 
(ii) a powerful $\mu$/e particle identification (PID), 
(iii) an improvement of the tracking pattern recognition,
(iv) a tracking independent trigger system and
(v)  an improved capability to remove background tracks mimicking the signal.
An extended description of the calorimeter requirements can be found in
Ref.~\cite{TDR}. In the following, 
the latest improvements on the calorimeter based algorithm for seeding of 
tracks are described. In Fig.~\ref{Fig:CALTRACK}, the distribution of hits in the Mu2e
detector is shown for single simulated CE events.
Most of the hits are due to background. By requiring the track hits to be 
in time with the most energetic cluster of the event,
in a time window of $\pm 50$ ns, the quantity of background hits is strongly reduced, thus improving the
relative reconstruction efficiency by $\sim $ 9\%.   Moreover, adding the
calorimeter time information to the reconstruction of the track, the
algorithm reduces hit ambiguities by 50 \%, resulting in a reduction
of a factor of 3 of the DIO background in the signal window.

Simulation studies show that, in order to meet the background rejection requirements, 
the calorimeter  should have: (i) a timing resolution  better than 500
ps , (ii) an energy resolution of $\mathcal{O}$(5\%) and (iii)  
a position reconstruction of $\mathcal{O}$(1 cm) to allow a correct match with the electron track
extrapolated to the calorimeter surface. 
A crystal based calorimeter is the adopted solution.

The crystal geometry has been optimized by means of a double disk setup 
to fulfill both the 
optimization of acceptance and 
the symmetric detection of e$^-$ and e$^+$. The distance between the disks (70 cm) and crystal
shape and dimension were tuned to increase acceptance. 
As a result, the baseline calorimeter  consists of 1650 square crystals  
arranged in two  disks having an inner radius of $\sim$ 35 cm  to
allow  low momentum DIO electrons to pass through the central hole,
without interacting with it. The calorimeter is 
located inside the DS: therefore it is requested to operate  in presence of  1 T axial
magnetic  field and in a vacuum of $10^{-4}$ Torr. It
should also be able to 
assure stable performance and efficiency in an environment where n, p and $\gamma$ backgrounds, 
from muon capture processes
and beam flash events, deliver a maximum dose  of $\sim$120 Gy/year
and a neutron fluence of up to 10$^{12}$ n/cm$^2$/year in the hottest
regions. The crystals intercept most of the neutron coming from
the target so that the fluence is strongly suppressed in the second
disk and in the 
rear part
of the first disk,  where a maximum 
fluence of up to 6$\times$10$^{10} $ n$_{\rm 1MeVeq}/$cm$^2$ has been estimated by our
simulations. 
 
\begin{figure}[!t]
\centering
\includegraphics[width=0.99\linewidth]{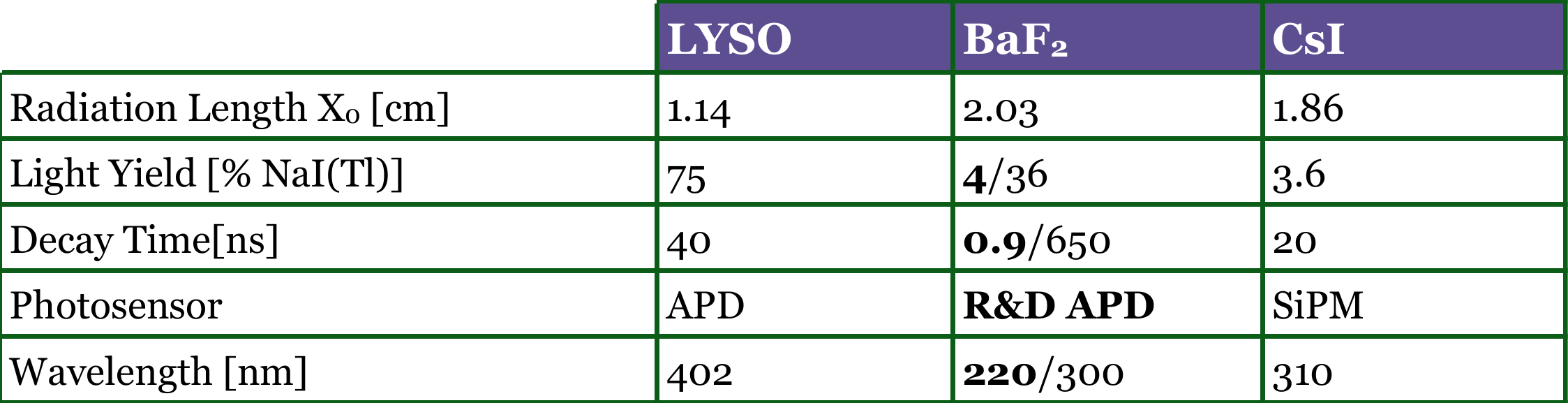}
\caption{Main properties of the considered crystals. }
\label{Fig:crystal_table}
\end{figure}

The first choice of calorimeter components consisted of LYSO crystals
\cite{NIM_LABIODOLA2012},  readout by two large area Hamamatsu avalanche 
photodiodes (APD) which granted the needed  characteristics 
but the sudden cost increase of these crystals, 
 in 2012-2013,  dictated the search 
for alternative crystal 
candidates. As a result, 
 BaF$_2$ crystal, a very fast scintillator, was chosen as our baseline, while 
retaining pure CsI as a further backup solution. 
In Fig.~\ref{Fig:crystal_table}, the comparison of properties among 
LYSO, BaF$_2$ and pure CsI is  shown. The BaF$_2$ is much faster than
LYSO but has a reduced light yield and emits at much smaller
wavelenghts. Its density is smaller than LYSO, but its
radiation length and Moliere radius 
are still acceptable  for Mu2e purposes.
The emission spectrum of the BaF$_2$ is composed of a fast (1 ns) component at
very short wavelength ($<$ 250 nm), and a larger slower (600 ns)
component at higher wavelengths ($>$ 250 nm).
The suppression 
of the ``slow'' component in the BaF$_2$ emission spectrum
is specifically discussed in subsec. \ref{photosensor}. The baseline
dimension of each crystal is $3.07\times3.07\times20$ cm$^3$, for a total
depth of 10 $X_0$, enough to contain the CE shower, considering that the CE 
impinges on the calorimeter surface with an average angle of 
$\sim$ 50$^{\circ}$.  The transverse dimension has been selected 
to optimize granularity and coupling to 
the photosensors while
fulfilling the requirements on position resolution 
and sufficient photoelectron collection.

A simulation of the baseline calorimeter geometry has been carried out
with Geant4 in the Mu2e simulation framework, by inserting all
details of shower development, pileup of background hits and most 
of the experimental effects such as: longitudinal response
uniformity along the axis, non-linearity on response and electronic noise. 
Scaling down from measurements in progress with PMTs (subsec.~\ref{baf2}), 
a light yield of 30 p.e./MeV  and an equivalent noise
of 300 keV per photosensor were assumed. With these inputs,
the simulation provides 
 an energy resolution at  the level of 4.5\% and a 
position resolution better than 7 mm for a CE. 
A signal shape simulation has not yet been  completed:
anyway a standalone simulation
indicates that a timing resolution better than 200 ps can be achieved.


\subsection{LYSO legacy}
A $5\times 5$ matrix prototype, built with ($30\times 30\times 200$) 
cm$^3$ LYSO  crystals and readout by means
of S8664-1010 APDs from Hamamatsu, was assembled in Frascati 
 to study the calorimeter performances.  The LYSO matrix has been 
 tested either with a tagged photon beam 
 and an
 electron beam~\cite{SIMONA}.
 A 4\%  energy resolution and
160 ps  timing resolution for 100 MeV incident particles were measured. 
These tests 
 allowed a comparison between the measured resolution and the MC simulation,
 so to check the MC reliability.

\subsection{Characterization of BaF$_2$ crystals}
\label{baf2}
\begin{figure}[!t]
\centering
\includegraphics[width=0.8 \linewidth]{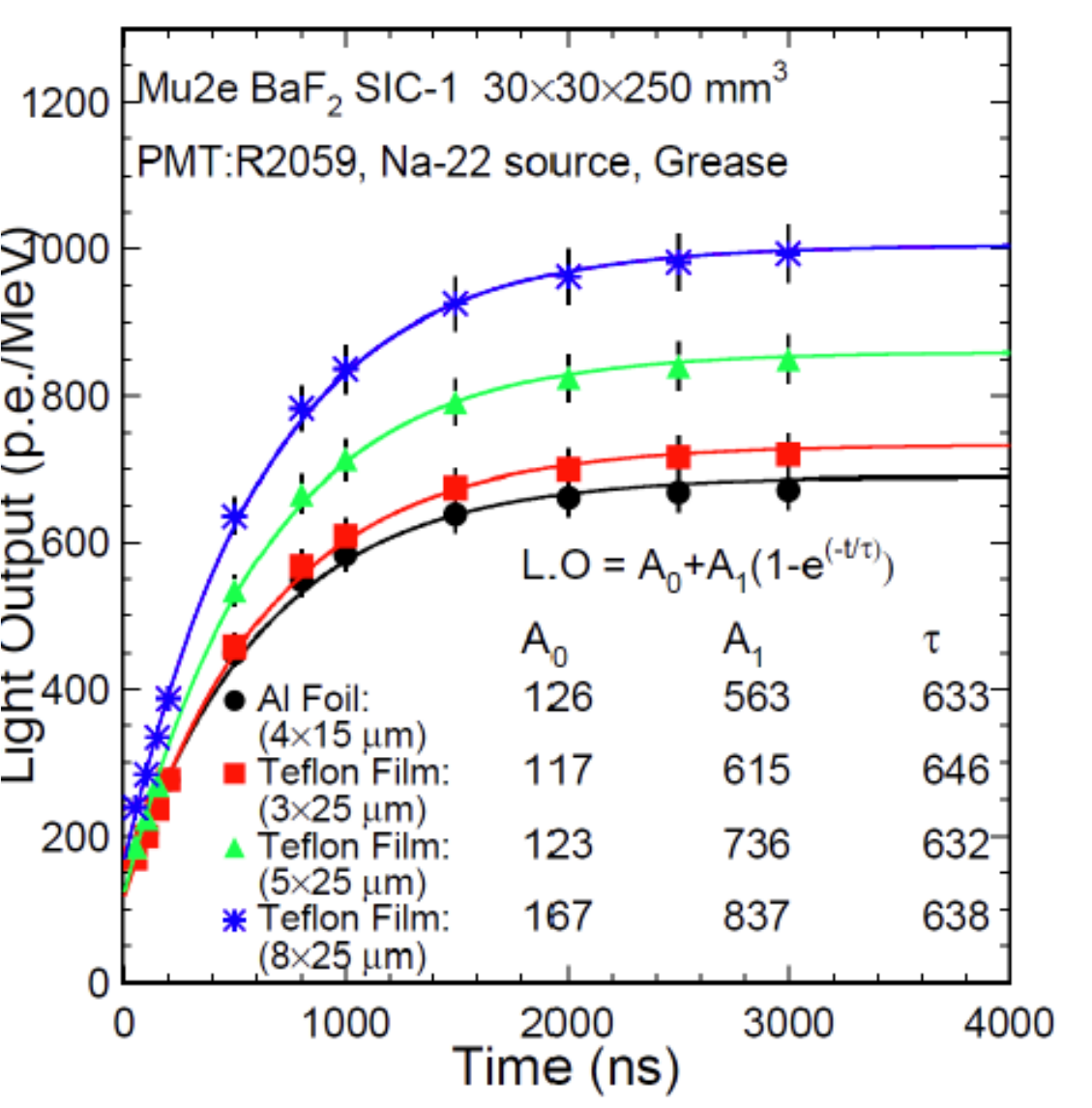}
\caption{Light output (L.O.) of a SICCAS BaF$_2$ crystal as a function
  of the ADC integration time. }
\label{Fig:Baf2}
\end{figure}

Light yield (LY) resolution and longitudinal
response uniformity (LRU),  of 26 prototype BaF$_2$ crystals have been measured:
20 crystals were produced by SICCAS (China), 2 from BGRI (China), 
and  2 from Incrom (Russia). The tests
were carried out by illuminating crystals along the axis with a
$^{22}$Na source. Crystals were readout by means of a UV extended photomultiplier.
 One of the 511 keV annihilation photons 
is used as calibration source, while the second one is used as tag to
clean the spectrum. 
In Fig.~\ref{Fig:Baf2}, the LY variation as a function of the
ADC integration time is shown for SICCAS crystals:
different colors corresponds to different wrapping (the highest
light yield is obtained with Teflon). The dependence of the response on
the integration time is well represented by a function of the form
$\rm{A_0}+\rm{A_1} (1-exp^{-t/\tau})$, where A$_0$, A$_1$ are the light yield for the fast and slow component
respectively and $\tau$ is the decay time of the slow component.
The longitudinal response uniformity has been studied by measuring the 
LY in 12 points along the axis: the distribution was fit with a linear function.
The SICCAS crystals show  a LY variation at both ends of 
$\pm$ 25\% with respect to the LY in the central position, that is slightly too large for our
purposes. BGRI and Incrom crystals show a smaller variation. 
The longitudinal transmittance of these crystals is $\sim$ 90\% at 220
nm, close to  the  theoretically calculated transmittance. 
Radiation damage tests are also in progress, both exposing crystals
to a ionization dose and to a neutron fluence. For BaF$_2$ the damage
due to the dose saturates at few hundreds Gy as reported elsewhere ~\cite{ZHU}. 
Light loss due to neutrons seems to be
contained at few \% for the fluence expected in Mu2e experiment lifetime.

\subsection{Photosensors R\&D}
\label{photosensor}
 The slow component of BaF$_2$ 
has to be suppressed due to the high event rate in the experiment.
A joint JPL/Caltech and RMD ~\cite{RMD} consortium has been established to
develop a new APD using 
a 9x9 mm$^2$ RMD device, adding
a superlattice microstructure to improve the UV sensitivity
and an atomic deposition layer to create an anti-reflection
filter rejecting the higher wavelengths (``solar blindness'').
A detailed description of the APD structure, the R\&D status 
and of the results obtained with
the first prototypes is reported
in~\cite{HITLIN}.

\subsection{Calorimeter Layout, FEE and readout}
\begin{figure}[!t]
\centering
\includegraphics[width=0.7\linewidth]{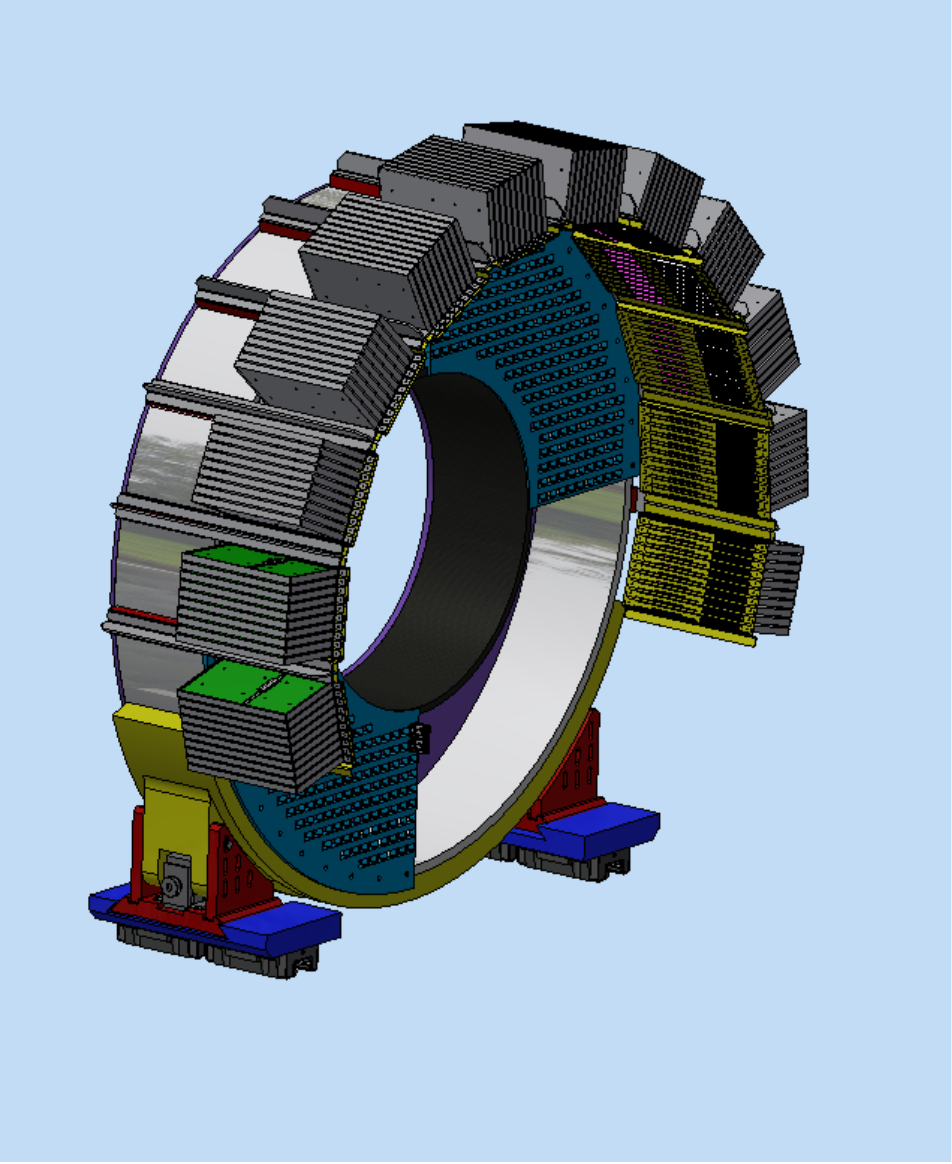}
\caption{CAD Drawing of one of the two disks. }
\label{Fig:mechanics}
\end{figure}

The calorimeter is composed of two disks. A preliminary CAD drawing of
one disk is shown in Fig.~\ref{Fig:mechanics}. The crystals will be
stacked inside an external stainless steal support cylinder. An inner
cylinder defines the hole where most of the low Pt particles will pass through.
This inner cylinder will be made of composite material and
will be connected to a front plate and to a back plate. The front face
plate will be integrated with the piping transporting the Fluorine
rich fluid used in the calorimeter calibration procedure (subsec.~\ref{calib}).
The external support cylinder will have feet allowing to
roll it over the insertion rails in the DS.  In
Fig.~\ref{Fig:mechanics} the back plate for supporting photonsensors
is also shown, as well as the arrangement of crates for the readout
electronics. Finite element analysis has been carried out for this
setup, showing negligible stress on the structure.

The electronics is composed by a FEE board directly connected to
photosensor. The FEE board amplifies signals and locally regulates the
bias voltages. A remote controller organized in a mezzanine board
distributes bias voltages to the APDs and LV power  to the
preamplifiers in group of 16 channels. The amplified signals,\
 via differential cables, reach 
a 16 channels board where are digitized.  
A sampling at 200 Msps with 12 bit resolution on
a dynamic range of 2 V allows to reach the required timing
performance and double pulse separation, while keeping the calorimeter
data throughput inside the DAQ operation range.

\subsection{Calorimeter Calibration system}
\label{calib}
\begin{figure}[!t]
\centering
\includegraphics[width=0.85\linewidth]{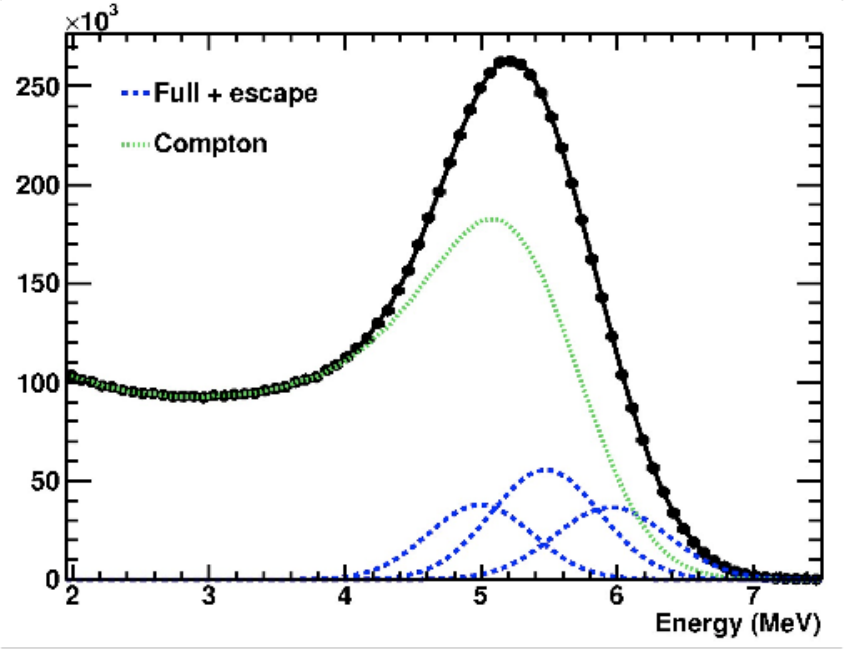}
\caption{Simulation of source calibration run. Black points: Reconstructed energy.
  Thick (thin) dashed line: contribution of photon peaks (Compton).}
\label{Fig:Calibration}
\end{figure}

The calorimeter calibration setup consists of: 
(i) a source system to
equalize the crystals response  and (ii)
a laser system to continuously monitor the fluctuations 
of the APD gains.  The source system is based on the BABAR scheme
\cite{BABAR_SOURCE}. 
A Fluorine rich fluid (Fluorinert) is irradiated by neutrons from a 
deuterium-tritium generator, and circulated into pipes to the front
face of the detector.
In Fig. \ref{Fig:Calibration}, the simulated energy spectrum of 
the  photons emitted by the excited Fluorinert is shown. 
It consists of a full annihilation peak at 6.13 MeV,
together with single and double escape peaks at 5.62 MeV and 5.11 MeV,
respectively, over the contribution of Compton photons. 
Even if single photon peaks cannot be separated, a clean
calibration signal appears over a rather flat background.  
In few minutes of running, 
the energy scale of each crystal can be set with high
precision ($< 0.5$\%). 
We plan to use it 
on regular base 
to follow the variation of the overall 
calorimeter response. Comparison with continuous laser runs will allow to
disentangle the contributions of possible fluctuations due to crystal response and photosensor
gain. 
A  prototype source system is being assembled in Caltech: an operation 
test is expect 
during the summer 2015.

\subsection{Backup Alternative}
\begin{figure}[!t]
\centering
\includegraphics[width=0.85 \linewidth]{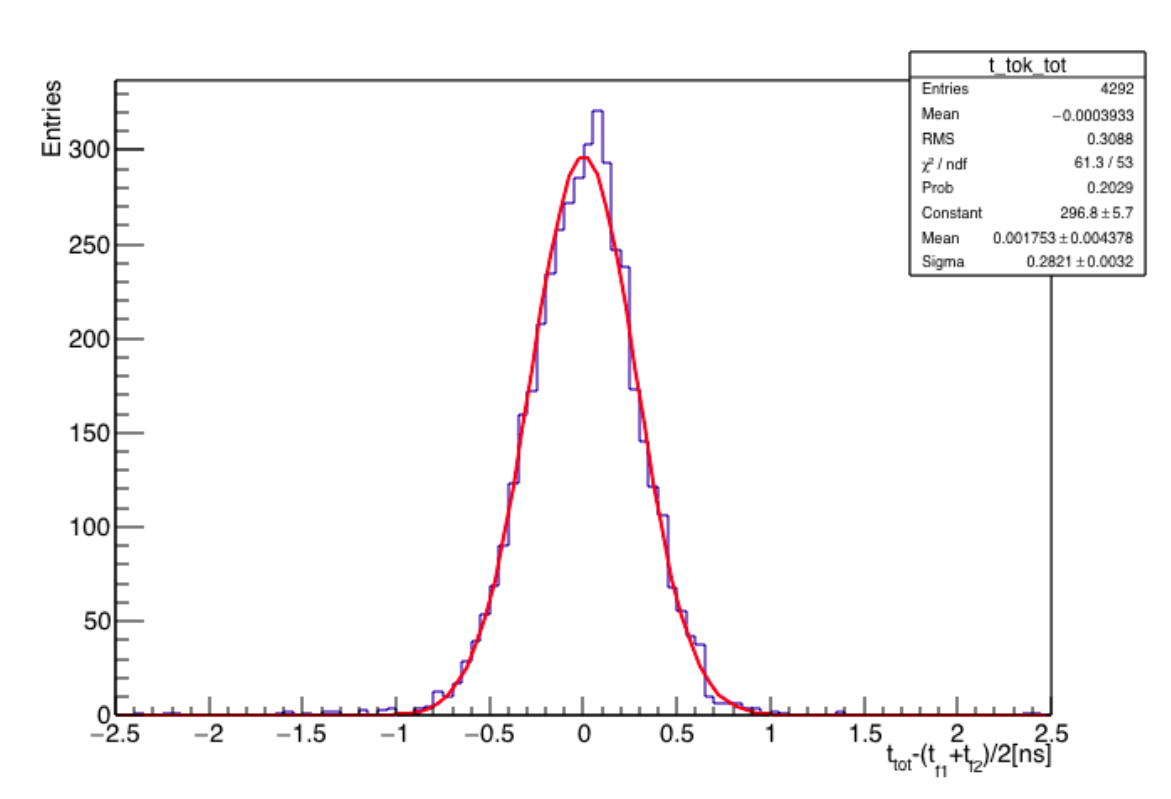}
\caption{Distribution of the time response of the CsI(pure)+MPPC
  matrix prototype to  100 MeV electrons impinging  at 50.$^{\circ}$ on the
  calorimeter surface.}
\label{Fig:TimeRes}
\end{figure}

To test 
a less expensive alternative to the BaF$_2$ 
baseline, 
pure CsI crystals,
read with the latest version of the UV extended Hamamatsu MPPC, 
have been studied.
The characteristics of 12 pure CsI square crystals, of 3x3x20
cm$^3$ dimension, were measured. 
The crystals were produced by three different vendors.
A detailed report of this study can be found in~\cite{Raffa}. 
In April 2015, 
10 new Silicon Protection Layer MPPC, with a
quantum efficiency (at 310 nm) improved by a factor of 4 with respect
to standard epoxy protected MPPC, were tested. 
 These photosensors  consist of an 
array of 16 3x3 mm$^2$ cells with 50 $\mu$m pixels, based on the new
Through Silicon Vias technology allowing reduced dark rates and
higher signal speed. A custom FEE board, to analogically sum all anode
signals, has been developed by INFN. A fast response and good timing
resolution ( $<$ 70 ps) has been observed,  when 
the sensor is illuminated with a fast blue
emitting laser light.  A $3\times 3$ CsI matrix 
has been built: each crystal was wrapped and was read out by one of the 
new photosensors optically connected to the crystal
by means of Bluesil past 7 optical grease. 
A short test beam has been carried out at
the test beam facility in Frascati National Laboratories 
to determine the response and resolution for 100 MeV
electrons. The following preliminary results have been achieved with
normal incidence electrons: (i) 7 \% energy resolution (ii)  250 ps
timing resolution without correcting for the trigger jitter. 
Also 
data with e$^-$ impinging at
an angle of $\sim$ 50$^{\circ}$ were collected and 
a 280 ps timing resolution
with a good Gaussian response was observed 
(Fig.~\ref{Fig:TimeRes}).
Optimization of the analysis is still under way: 
anyway the presented results already
satisfy the calorimeter required specifications.

\section{Conclusions}
The Mu2e calorimeter will provide complementary information to the
tracker system and will be used for PID, seeding
of tracks, triggering and validating the CE track candidates. 
The baseline system consists of two disks
of 1650 BaF$_2$ square crystals, each one readout by two ``solar blind''
APDs. Simulated performances match requirements 
and  the test of the first  APD prototypes is under way. A 
backup alternative, based on pure CsI crystals
read by UV extended MPPC, is also under test and the preliminary 
results prove to 
match the requirements.
An irradiation test of all components is under way to certify their
viability in the environmental condition of the experiment.
A technology review is set for summer 2015 to 
confirm the validity of the baseline choice and allow to 
complete the engineering design. 



\begin{thebibliography}{10}
\expandafter\ifx\csname url\endcsname\relax
  \def\url#1{\texttt{#1}}\fi
\expandafter\ifx\csname urlprefix\endcsname\relax\def\urlprefix{URL}\fi

\bibitem{TDR}
L.~Bartoszek et al.\ (Mu2e Experiment), ``The Mu2e Technical Report'',
arXiv:1501.05241 (2015).

\bibitem{SINDRUM} W.Bertl et al., (Sindrum-II), `` A search for muon
  to electron conversion in muonic gold'', Eur. Phys. J. C47, 337 (2006).

\bibitem{NIM_LABIODOLA2012}
J.~Budagov et al.,
``The calorimeter project for the Mu2e experiment'', 
Nucl.~Instrum.~Meth.~A 718 (2013) 56.

\bibitem{SIMONA} N. Atanov et al.,``Energy and timing resolution for
  a LYSO matrix prototype of the Mu2e experiment'', these proceedings.

\bibitem{ZHU} R.Y. Zhu et al.,``The next generation of crystal detectors'',
  these proceedings.
  
\bibitem{RMD} http://rmdinc.com/avalanche-photo-diodes/.

\bibitem{HITLIN} D.G. Hitlin et al., ``An APD for the detection of the
  fast scintillating component of BaF$_2$'',  these proceedings.

\bibitem{BABAR_SOURCE} B.Aubert et al., ``The BABAR detector'',
  Nucl.~Instrum.~Meth.~A479 (2002) 1.


\bibitem{Raffa} M.Angelucci et al., ``Longitudinal uniformity, time
  performances and irradiation test of pure CsI crystals'', these proceedings.



\end{thebibliography}
\end{document}